# Intrinsic Optical Transition Energies in Carbon Nanotubes

Yan Yin,[1] Stephen Cronin,[2] Andrew Walsh,[1] Alexander Stolyarov,[2] Michael Tinkham,[2] Anthony Vamivakas,[3] Wolfgang Bacsa,[3,4] M. Selim Ünlü,[3,1] Bennett B Goldberg,[1,3] and Anna K. Swan[3,*]

[1]*Physics department, Boston University, Boston, MA 02215,*

[2]*Physics department, Harvard University, Cambridge, MA 02138,*

[3]*Electrical and Computer Engineering Department, Boston University, Boston, MA 02215*

[4]*LPST-IRSAMC CNRS, Université, Paul Sabatier, 31062 Toulouse, France*

Intrinsic optical transition energies for isolated and individual single wall carbon nanotubes grown over trenches are measured using tunable resonant Raman scattering. Previously measured $E_{22}^S$ optical transitions from nanotubes in surfactants are blue shifted 70-90 meV with respect to our measurements of nanotubes in air. This large shift in the exciton energy is attributed to a larger change of the exciton binding energy than the band-gap renormalization as the surrounding dielectric constant $\varepsilon$ increases.

After a decade of intense study, fluorescence was observed in single wall carbon nanotubes, where the key advancement was to isolate the tubes in surfactant micelles [1]. Shortly thereafter photoluminescence excitation (PLE) maps of the same system helped to identify the chiral index (n,m) of a tube by associating the characteristic distribution of the PLE resonances peaks ($E_{22}^S$, $E_{11}^S$) with predicted transition energies from tight binding theory [2]. The $E_{11}$ transition energies were higher than predicted and the ratio $E_{22}/E_{11} \cong 1.7$ rather than 2. Theoretical work predicts that the Coulomb driven exchange interaction gives rise to a large increase of the band gap energy counteracted by a somewhat smaller exciton binding energy shift [3,4]. The resulting energy levels are predicted to be higher [3-7] than those expected from tight binding calculations in accordance with PLE measurements [2].

Recently, similar mappings of $E_{22}^S$ vs. diameter were achieved for single wall nanotubes in surfactant solution using tunable resonant Raman scattering (RRS). In RRS the excitation wavelength is tuned through the optical absorption while monitoring the Stokes and Anti-Stokes scattering intensity of a Raman active mode. The RRS measurements of carbon nanotubes in solution monitored the radial breathing mode (RBM) intensity, where the RBM frequency is inversely proportional to the tube diameter, yielding an optical transition energy vs. diameter map [8,9]. The RRS data confirmed the PLE tube assignments and demonstrated that the $E_{22}^S$ measured by the two techniques are indeed the same within a maximum deviation of 35 meV [9].

All of the atoms in a single-wall nanotube (SWNT) are external, and hence, optical transitions can be strongly influenced by the environment. For example, tubes dispersed with different cationic, anionic and nonionic surfactant molecules show variations in the optical resonances up to ~25 meV [10]. PLE from tubes suspended in air between pillars show a shift in $E_{22}^S$ and $E_{11}^S$ compared to tubes wrapped in surfactants [11]. Any comparison with theory of the intrinsic electronic properties of SWNT's requires probing of isolated and individual tubes unaffected by the environment. Our aim in this work is to probe the intrinsic optical properties of nanotubes by examining individual nanotubes suspended in air.

We investigate a series of individual SWNTs with different diameters and chiralities, suspended over $1-2\mu m$ wide trenches and record in detail their resonant Raman profiles. We observe narrow resonances from $E_{22}^S$ transitions. The resonances show a systematic shift to lower energies by 70-90 meV compared to individual SWNTs in sodium dodecyl sulfate (SDS) solution. We ascribe the systematic down-shift compared to tubes in SDS to the change in the surrounding dielectric medium.

The samples are prepared by first etching $1-2\mu m$ wide trenches with fiduciary markers in quartz substrates to make it possible to locate a specific nanotube repeatedly. The SWNTs are grown over the trenches by chemical vapor deposition [12]. A tunable Ti-sapphire laser is used for Raman excitation in the range 720-830 nm. Laser line rejection is achieved by tilt-tuning of filters with matching glass slabs to compensate for beam offset. The use of filters and a single grating offers a high through-put system enabling single tube detection. The laser beam is focused by a 100X objective with the Gaussian spot-profile $FWHM = 0.47 \mu m$ at $E_{laser} = 785 nm$ with 1-2 mW constant excitation laser power during a resonance profile measurement. Direct measurements of Stokes and anti-Stokes intensity ratios show that no heating of the nanotubes takes place under such powers. The laser beam is scanned along a $77 \mu m$ long trench on the sample and typically 3-10 resonant tubes are found. (Stokes RBM peak count rates are 30-350 counts/second). Each resonant excitation profile (REP) is measured twice with staggered 2 nm separation in excitation wavelength to ensure repeatability.

Figure 1 shows the resonant Raman Stokes and anti-Stokes RBM REPs from an individual isolated tube suspended in air. This tube has $RBM = 258 cm^{-1}$, and is assigned as a (9,4) tube. Figure 1 a) shows the raw 2D spectral data map after subtracting a linear background. The anti-Stokes (AS, left) and Stokes (S, right) resonances are both clearly observed, and their intensity maxima are shifted in excitation energy by the RBM phonon energy due to the resonant enhancement for both the incoming and scattered light [13]. The resulting resonant energy profiles from the AS and S RBM peaks are shown in figure 1 b). The Stokes resonance profile is fitted using the time-dependent, third-order perturbation formulation for Raman scattering in a one dimensional (1D) system [14]. We can use either a delta function representing an exciton level, or the well known band-edge van Hove singularity density of states (DOS) for 1D. While both approaches produce a symmetric resonance profile, here we use the van Hove singularity DOS [15] to fit the Stokes REP to compare our results with earlier studies [8,9]. In this case the Raman intensity as a function of excitation laser photon energy $I(E_l)$ can be written as [14]:

$$I(E_l) \propto \beta_{S,AS} \times \left| \frac{1}{\sqrt{E_l - E_{ii} - i\eta}} - \frac{1}{\sqrt{E_l \mp E_{ph} - E_{ii} - i\eta}} \right|^2 \quad\quad 1)$$

The broadening parameter η is added to account for the finite lifetimes of the intermediate states in the scattering process. The factor $\beta_{S,AS}$ accounts for the phonon occupancy ratio with $\beta_{S,AS} = \exp[-E_{phonon}/kT]$ for anti-Stokes and $\beta_{S,AS} = 1$ for Stokes scattering. The fit of the Stokes REP shown in Fig 1b) yields $E_{22}^S = 1.629 eV \pm 1.5 meV$ and $\eta = 17.8 \pm 3.5 meV$. The same parameters are used to calculate the anti-Stokes scattering profile, after changing sign for the phonon energy and using $\beta_{S,AS}$. The resulting curve is plotted in figure 1 b). Equation 1 not only fits the Stokes data well, but the calculated AS profile also matches the AS data with no adjustable parameters. This demonstrates that the nanotube remains at room temperature (300K), and that the resonance energy and broadening parameters can be determined accurately. RRS measurements of nanotubes in dry nitrogen atmosphere, before and after heating, exhibit the same resonance energy as nanotubes in air [16]. Hence, we see no trace of water adsorbed on the nanotubes in air, probably due to the hydrophobic nature of graphite. Repeated measurements on the same nanotube on different occasions gave the same resonance energy within a few meV.

The line shape, broadening $\eta$ and $E_{22}^S$ of the RBM resonance excitation profiles were measured on 18 different isolated, individual semiconducting tubes as well as 7 metallic tubes. The symmetry of each REP line shape was determined by calculating a symmetry ratio, $R = I_{above}/I_{below}$, integrating the spectral weight below and above the center of the fitted peak, shown in Fig. 1c). As expected, within experimental uncertainty, the measured RBM REP line shapes are symmetric for the $E_{22}^S$ and also for the $E_{11}^M$ metallic resonances [17].

A histogram of the broadening parameters $\eta$ for all measured nanotubes is shown in Fig. 1d). The minimum $\eta$ value we observed is 8.8 meV, similar to the only previously reported REP measurement of one single tube on a Si substrate [13]. Half of the measured nanotubes have a broadening factor <19 meV, with a decreasing number of nanotubes with larger $\eta$. The few tubes with larger broadening parameters we believe are affected by environmental perturbations or defects. Therefore we ascribe the narrow broadening ($\cong 10 meV$) as indicative of either the intrinsic broadening of a single suspended tube, or broadening limited by the finite suspension length (~1-2 $\mu m$) across the trench. It is interesting to note that the line-width of the individual Raman RBM modes (typically 4-11 cm$^{-1}$) is not correlated with the REP broadening $\eta$. Hence, the Raman RBM line-width is not sensitive to perturbations that broaden the resonant energy profile of a nanotube and cannot be used as an indicator of the nanotubes electronic coupling to the environment. The average η is ~14 meV for isolated tubes suspended in air whereas RRS measured on HiPCO tube ensembles wrapped in SDS are broadened by 65 meV [9] and on bundles by 120 meV [9]. The larger line widths of the REP profiles in SDS and bundles raises the question if these larger broadening factors are due to interactions with the environment or simply are due to inhomogeneous broadening.

PLE measurements from a single nanotube also suspended in air show a symmetric $E_{22}^S$ excitation profile with a width of 44 meV, ~4-5 times broader than our REP data [18]. This difference can be explained by the difference in the Raman and PL process: In PL, absorption, intra- and inter-band phonon relaxation and emission rates all involve real states, where transition rates can be calculated using Fermi's golden rule [19]. On the other hand, resonance Raman scattering is described by third order time dependent perturbation theory with a three-step phase coherent quantum mechanical process resulting in scattered Raman signal. This process includes a virtual state and does not allow electron scattering between real states in either the incoming or outgoing resonance situation. Hence, there is no reason to expect the same broadening in these two cases.

Figure 2 shows a map of the measured optical transitions $E_{22}^S$ versus the RBM frequencies $\omega_{RBM}$. The error in fitting $E_{22}^S$ is smaller than the size of the symbols used. Each resonance is shown with an energy range that indicates the broadening $\eta$ measured for that particular tube. The tubes with the narrowest line-broadening, $\eta <$ 19 meV (red symbols) will be discussed primarily [18]. For comparison, the optical transition values from nanotubes in SDS solutions are also shown with open square symbols [9].

The inverse relationship between RBM frequency and diameter, $\omega_{RBM}(cm^{-1}) = A/dia(nm) + B$, is used to assign specific chiral indices (n,m) to the resonances. In SDS solution the relationship is $\omega_{RBM}(cm^{-1}) = 223/dia(nm) + 10$ and the (n,m) assignment is anchored by the geometric family pattern in the $E_{ii}$, versus $\omega_{RBM}$ map [9,20]. We use the group of 8 nanotubes in the range $255-275 cm^{-1}$ as our geometric anchor. The negative curvature of that branch makes it clear it belongs to a $(n-m) \mod 3 = 2$ branch and the RBM frequencies place it in family 22 (2n+m). Since the observed branch has the same slope and same RBMs as the (9,4), (10,2) and (11,0) tubes in the SDS family 22 branch [8,9], we assign the same (n,m) values to these SWNTs and conclude that nanotubes suspended in air have similar constants A and B relating $\omega_{RBM}$ and diameter as nanotubes in SDS solution [21].

The $E_{22}^S$ energies for the tubes in family 22 at $255-275 cm^{-1}$ are ~90 meV lower than SWNTs in SDS solution. Different tubes of the same (n,m) in family 22 show an energy spread of ~ 6-7 meV, close to our precision ($<\pm 4 meV$) in measuring $E_{22}$. The two tubes around $205 cm^{-1}$ with narrow line widths are assigned as (14,1) nanotubes in family 29 with the $E_{22}^S$ energy $\sim 70 \pm 7 meV$ lower than nanotubes in SDS. Hence, the $E_{22}^S$ energies of nanotubes in air are 70-90 meV lower in energy than for nanotubes in SDS solution. Seven measured metallic $E_{11}^M$ resonances are also shown together with an empirical extrapolation of expected metallic optical resonance positions for nanotubes in a micelle solution [22].

In the following we argue that PLE ensemble measurement study on nanotubes suspended in air by Lefebvre *et al* [11] should have a different (n,m) assignment, where the new assignment yields energy shifts consistent with our findings. RRS maps of resonant nanotubes uses the RBM frequencies to anchor (n,m) assignment. Assignments of PLE peaks to specific nanotubes require a large map for an unambiguous assignment [2]. For the PLE study of the suspended nanotubes in air, the $E_{11}$, $E_{22}$ peaks were equated with the closest SDS peaks, which yielded an apparent average blue-shift of $E_{11}$ (~28 meV) and $E_{22}$ (~16 meV). The work of Telg *et al.* [8] and Fantini [9] *et al.* shows that the $E_{22}$ resonances are the same for RRS and PLE measurements. Our measured $E_{22}$ energies in family 22 coincide with the measured PLE $E_{22}$ absorption energies for the three ($E_{11}$, $E_{22}$) peaks identified as belonging to family 25, also a branch $(n-m) \mod 3 = 2$. Based on the curvature of the three PLE peaks of $(n-m) \mod 3 = 2$, the Lefebvre data is most likely missing the zigzag tube. Our (9,4) tubes ($E_{22}$=1.630 eV) matches the $E_{22}$ of ($E_{11}$=1.011, $E_{22}$=1.601) and our (10, 2) tubes ($E_{22}$=1.594 eV) matches the $E_{22}$ of ($E_{11}$=1.060, $E_{22}$=1.593). With this re-assignment, the PLE study of tubes suspended in air instead show a downshift of both $E_{11}$ (115 and 117 meV) and $E_{22}$ (110 and 90 meV) rather than the previously assigned blue-shift.

Compared to our experimental results for CNTs in air, CNTs suspended in SDS solution show a 70meV-90meV blue shift. This blue shift, associated with a higher dielectric constant environment, is evidence of exciton presence in CNTs. The optical excitation energy (the exciton energy) can be divided into a non-interacting single electron contribution and many-body effects [4, 7, 23, 24],

$E_{ii}(\varepsilon) = E_{ii}^{SP} + E_{ii}^{BGR}(\varepsilon) - E_{ii}^{XB}(\varepsilon)$, where $E_{ii}^{BGR}(\varepsilon)$ stands for the band-gap renormalization contribution and $E_{ii}^{XB}(\varepsilon)$ is the exciton binding energy, both of which depend on the external dielectric constant $\varepsilon$. For an exciton in a higher dielectric environment, the exciton binding energy $E_{ii}^{XB}(\varepsilon)$ will be screened to be smaller than the case in a low dielectric environment [24]. This exciton screening effect will push optical transition energies higher for a CNT in a higher dielectric environment.

On the other hand, screening of the $E_{ii}^{BGR}(\varepsilon)$ in a higher dielectric environment will decrease the optical transition energy. The lower values of $E_{22}$ in our result show that the change of exciton binding energy $\Delta E_{22}^{XB}$ is larger than the change of band-gap renormalization $\Delta E_{22}^{BGR}$ when $\varepsilon$ goes from 1 in air to a higher value in SDS solution. The calculations by Ando and co-workers predict that the value of exciton binding energy $E_{ii}^{XB}$ is smaller than band-gap renormalization $E_{ii}^{BGR}$ for $\varepsilon=1$ [4]. However, when the dielectric constant increases, the resulting energy shift direction is decided not only by the values of $E_{22}^{XB}$ and $E_{22}^{BGR}$ at $\varepsilon=1$, but also how fast $E_{22}^{XB}$ and $E_{22}^{BGR}$ decrease as a function of $\varepsilon$ [7, 23, 24]. If $E_{22}^{XB}$ has a stronger functional dependence on $\varepsilon$ than $E_{22}^{BGR}$, as expected [7,24], the change $\Delta E_{22}^{XB}$ can be larger than the change $\Delta E_{22}^{BGR}$, even with $E_{22}^{BGR} > E_{22}^{XB}$ as predicted by Ando [3,4]. Hence, we interpret the blue-shift in energy with increased dielectric environment as due to a stronger screening of $E_{22}^{XB}(\varepsilon)$ than $E_{22}^{BGR}(\varepsilon)$ with increasing $\varepsilon$.

In summary, we have measured the resonant Raman excitation profiles on individual and isolated semiconducting single wall carbon nanotubes suspended in air. The measured broadening ~10 meV for an individual, suspended SWNT is significantly lower than for nanotubes in solution and is associated with the intrinsic lifetime broadening for the intermediate states. Optical transition energies are found to be 70-90 meV lower than for nanotubes wrapped in SDS. The energy difference is attributed to the difference in the dielectric constant for the two measurements. This result provides evidence of excitons and shows that a change in the external dielectric environment from $\varepsilon = 1$ in air to higher $\varepsilon$ in SDS solution will result in a larger change of the exciton binding energy than the change of the electron-self energy.

The authors acknowledge helpful discussions with Antonio Castro-Neto, Francisco Guinea and Millie Dresselhaus. This work is supported in part by NSF # NIRT ECS-02-0210752 and a Boston University SPRInG grant.

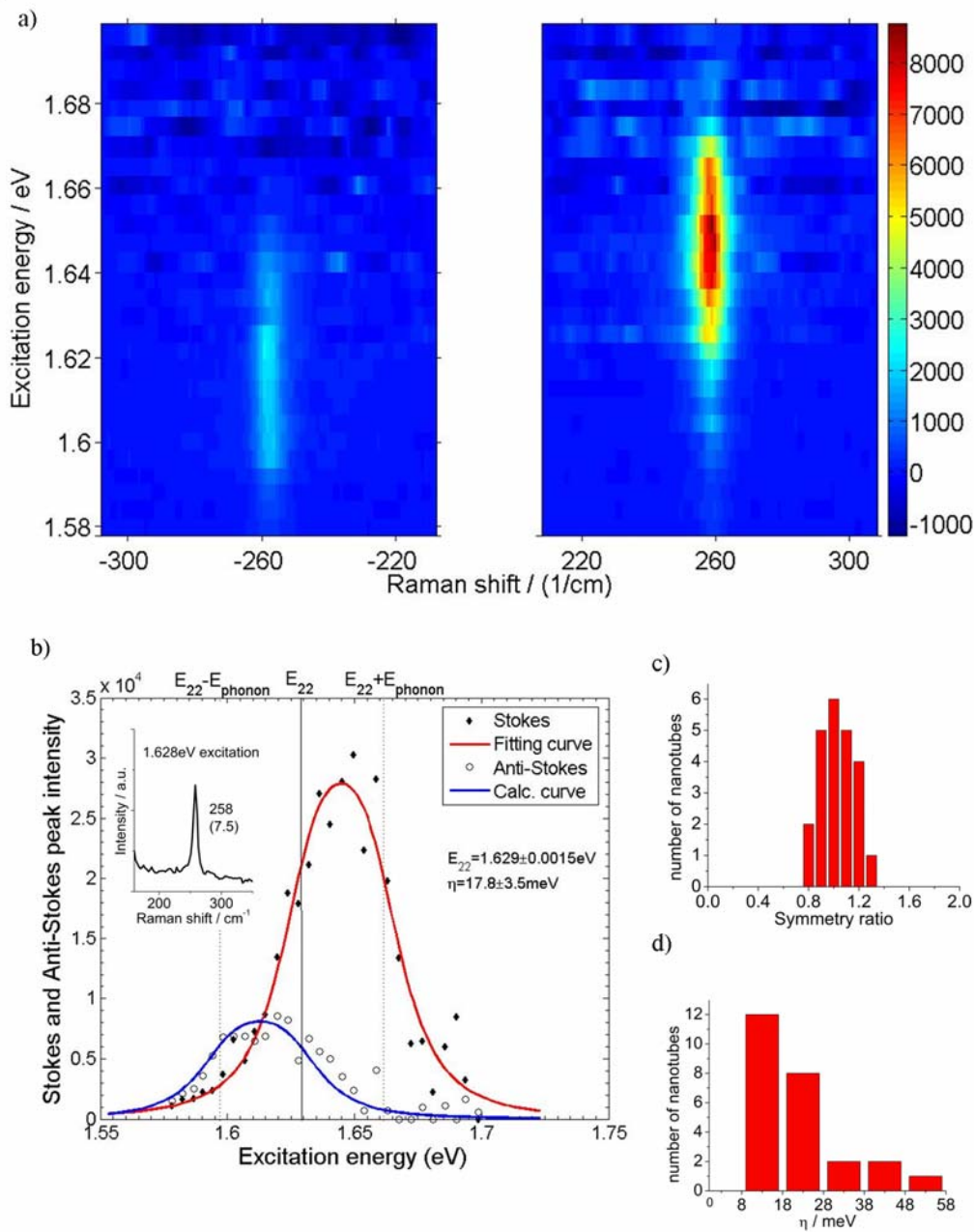

Fig 1. a) 2D intensity plot of anti-Stokes (left) and Stokes (right) RBM spectra as a function of excitation energy for one individual suspended SWNT. The RBM mode is $258 cm^{-1}$. b) Stokes and anti-Stokes excitation profiles from integration of the RBM peak intensity at each excitation energy. The red line shows the Stokes REP fitted by Equation 1, while the blue line is the calculated REP using the identical energy and broadening parameters plus a 300K phonon bath, showing excellent agreement. The vertical lines show the incoming (solid) and scattered (dotted) resonances. The inset shows the Raman RBM spectrum for $E_{laser} = 1.65 eV$. c) Histogram of the symmetry ratio for 23 tubes for the measured Stokes peaks, average of $R = 1.02 \pm 0.14$. d) Histogram of the 25 measured broadening parameters $\eta$.

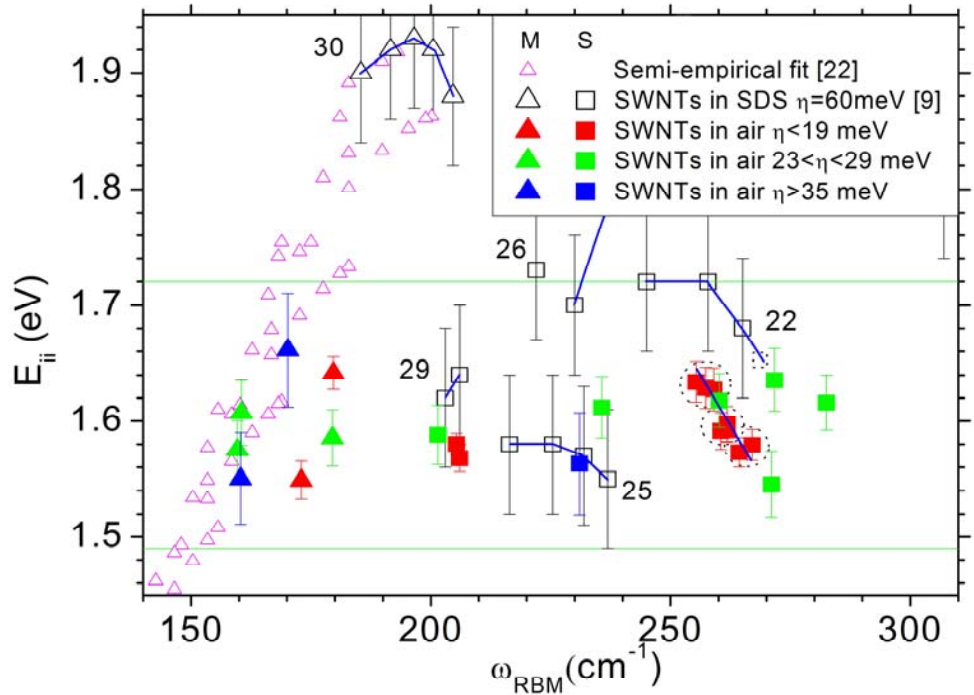

Fig 2. Experimental plot of $E_{ii}$ vs. $\omega_{RBM}$ for the 18 semiconducting SWNTs (filled squares) and 7 metallic SWNTs (filled triangles) measured by tunable RRS in air, and for comparison, RRS in SDS [9] and a semi-empirical fit for metallic SWNTs [22]. The numbers denote the 2n+m families. The energy ranges shown for each point are the experimentally measured broadening factors $\eta$. The horizontal lines indicate our experimentally measurable range for excitation photon energy. The overall width of the resonance window for an RBM REP can be estimated by $\eta + E_{phonon} \cong 25\text{-}50 meV$.

---

[*] Electronic address: swan@bu.edu